\documentclass[12pt]{article}
\usepackage{amsmath}
\usepackage{amsthm, amssymb}
\usepackage{multirow, booktabs}
\usepackage{natbib}
\usepackage{bm}
\usepackage{pdfpages}
\usepackage{geometry, setspace}
\allowdisplaybreaks[4]
\DeclareMathOperator\var{var}

\def\IF{\mathrm{IF}}

\def\expit{\mathrm{expit}}
\newtheorem{assumption}{Assumption}
\newtheorem{lemma}{Lemma}
\newtheorem{theorem}{Theorem}
\newcommand\independent{\protect\mathpalette{\protect\independenT}{\perp}}
\def\independenT#1#2{\mathrel{\rlap{$#1#2$}\mkern2mu{#1#2}}}

\title{\bf Adjusted Nelson--Aalen estimators by inverse treatment probability weighting with an estimated propensity score}

\author{Yuhao Deng$^1$ and Rui Wang$^2$ \\
{\small 1 University of Michigan \ 2 University of Washington}}

\begin{document}

\maketitle

\begin{abstract}
{Inverse probability of treatment weighting (IPW) has been well applied in causal inference to estimate population-level estimands from observational studies. For time-to-event outcomes, the failure time distribution can be estimated by estimating the cumulative hazard in the presence of random right censoring. IPW can be performed by weighting the event counting process and at-risk process by the inverse treatment probability, resulting in an adjusted Nelson--Aalen estimator for the population-level counterfactual cumulative incidence function. We consider the adjusted Nelson--Aalen estimator with an estimated propensity score in the competing risks setting. When the estimated propensity score is regular and asymptotically linear, we derive the influence functions for the counterfactual cumulative hazard and cumulative incidence. Then we establish the asymptotic properties for the estimators. We show that the uncertainty in the estimated propensity score contributes to an additional variation in the estimators. However, through simulation and real-data application, we find that such an additional variation is usually small.} 
\end{abstract}

\section{Introduction} \label{sec1}

For time-to-event data analysis, a key problem is that some failure times are unobserved due to right censoring. In the presence of competing risks, the cumulative incidence function (CIF) of an event is usually adopted as the estimand, which is always well defined no matter whether this event would eventually happen \citep{prentice1978analysis}. We may have two strategies to estimate the population-level CIF: the first is to directly estimate the incidence by weighting the observed event counting process by the inverse of the uncensored probability, and the second is to estimate the cause-specific hazards and then transform the hazards to cumulative incidence. To minimize the models imposed, the analysis for time-to-event data usually proceeds on the scale of hazards by assuming random censoring. Therefore, identifying the cause-specific hazards is the central aim in the presence of competing risks \citep{lau2009competing}. 

Indeed, it is challenging to specify an appropriate hazard model when there is no information on how covariates influence the hazards \citep{logan2008comparing}. Modeling the post-treatment time-varying covariates is both statistically and computationally challenging, which limits the practical use \citep{rytgaard2022continuous}. Although model-based methods (for example, Cox regression) have been well applied to estimate the coefficients of risk factors, it is not straightforward to infer the population-level cumulative incidence where the distribution of baseline covariates should be averaged out. Nonparametric approaches are sometimes preferred since there is no model restriction on the hazards, especially in randomized trials. Nonparametric estimation proceeds based on the counting processes of events, such as the Kaplan--Meier estimator using product limits and Nelson--Aalen estimator transforming the cumulative hazard \citep{kaplan1958nonparametric, nelson1972theory, aalen1978nonparametric}. At a tolerable cost of efficiency loss, the nonparametric approaches give consistent estimators for the cumulative incidence (or survival function) in finite sample.

A scientific question is to estimate the population-level estimand (cumulative incidence function) and average treatment effect from observational studies. In observational studies, the covariates may not be identically distributed between treatment groups. Inverse probability weighting (IPW) has been well applied in causal inference to deal with imbalanced covariates \citep{rosenbaum1983central, robins2000marginal}. An unbiased estimator for the population-level estimand is obtained by weighting the individuals in a trial using the inverse of the treatment probability. The propensity score summarizes the information of baseline covariates into a one-dimensional statistic. Conditioning on the propensity score yields balanced covariate distributions between treated and control groups. The idea of IPW can be extended to time-to-event outcomes with random right censoring. By weighting the counting processes with the inverse treatment probability, the adjusted Nelson--Aalen estimator (or asymptotically equivalently, weighted Kaplan--Meier estimator when there is only one terminal event) and weighted Aalen--Johansen estimator can be constructed \citep{winnett2002adjusted, xie2005adjusted, austin2014use, mao2018propensity, hu2020modified, vakulenko2023causalcmprsk}. The variance of the resulting estimator can be explicitly derived using the martingale theory if the treatment propensity score is known.

However, two problems are left unstudied in using IPW for time-to-event data. First, the nonparametric Nelson--Aalen and Aalen--Johansen estimators assume homogeneity of the hazard functions, in that all individuals in a single treatment group share the same risk of failure events \citep{nelson1972theory, aalen1978nonparametric, aalen1978empirical}. No covariates are incorporated in the models. Second, existing work did not consider the uncertainty of the estimated propensity score \citep{cole2004adjusted, austin2014use}. It is well known that using the estimated propensity score in the IPW estimator may lead to a slightly different standard deviation compared to using the true propensity score \citep{hirano2003efficient}. However, it is unknown how the uncertainty of the estimated propensity score will affect the inference regarding the adjusted Nelson--Aalen and Aalen--Johansen estimators.

In this article, we formalize the adjusted Nelson--Aalen estimator by inverse treatment probability weighting in the competing risks setting. We focus on the Nelson--Aalen-type estimator rather than the Aalen--Johansen-type estimator to avoid product limits in the estimators. We show the identifiability of the counterfactual cumulative incidence function for a competing event. If the propensity score is known, we construct a martingale from the counting processes, and derive the unbiasedness and the finite-sample variance of the estimated counterfactual cause-specific hazard function. If the propensity score is unknown but the estimated propensity score is regular and asymptotically linear, we derive the influence function of the estimated counterfactual cause-specific hazard. The asymptotic property of the estimated counterfactual cumulative incidence function is then established using the empirical process theory. We find that omitting the uncertainty of the estimated propensity score may lead to a biased variance estimator. However, simulation studies show that the empirical bias is usually tiny.

The remainder of this article is organized as follows. In Section \ref{sec2}, we list the assumptions for identifying the counterfactual cumulative hazards and incidence functions, and then give the adjusted Nelson--Aalen estimator by inverse probability of treatment weighting. In Section \ref{sec3}, we establish the asymptotic properties for the estimator, when the propensity score is known and when the propensity score is estimated, respectively. In Section \ref{sec4}, we conduct simulation studies to assess the influence of using an estimated propensity score in the adjusted Nelson--Aalen estimator. We also perform a sensitivity analysis using a misspecified propensity score in the Supplementary Material. In Section \ref{sec5}, we apply the proposed estimator to study the effect of transplant modalities on relapse and non-relapse mortality. Finally, we end this article by pointing out some extensions in Section \ref{sec6}.

\section{Estimation} \label{sec2}

Let $X$ be the baseline covariates. Let $A$ be the treatment indicator, where $A=1$ stands for the active treatment and $A=0$ stands for the control (placebo). If there is only one terminal event, we let $\tilde{T}^a$ be the potential time to the event under the treatment condition $A=a$. If the event does not happen at the end of study $t^*$, we can denote $\tilde{T}^a > t^*$ since we are not interested in the occurrence of events after $t^*$. With competing events, the time to an event may not be well defined. It is more convenient to introduce the event counting processes. Under the stable unit treatment value assumption (SUTVA), let $\tilde{T}^a$ be the potential time to the first event under the treatment condition $A=a$ and let $\tilde\Delta^a \in \{1,\ldots,J\}$ be the potential event indicator. We denote $\tilde{T}_j^a = \tilde{T}^a$ if $\tilde\Delta^a=j$ and $\tilde{T}_j^a = \infty$ otherwise. Let $\tilde{N}_j^a(t) = I\{\tilde{T}_j^a\leq t\}$ be the potential event counting process for the event $j$ and $\tilde{Y}^a(t) = I\{\tilde{T}^a \geq t\}$ be the potential at-risk process. All competing events share the same at-risk process.

Subject to right censoring, we cannot fully observe the event counting processes. Let $\tilde{C}^a$ be the potential censoring time under the treatment condition $A=a$, then the potential follow-up time is $T^a = \tilde{T}^a \wedge \tilde{C}^a$. Let $\Delta^a = \tilde\Delta^a I\{\tilde{C}^a\geq\tilde{T}^a\}$ be the event indicator, so that $\Delta^a=0$ if censoring happens first and $\Delta^a=\tilde\Delta^a$ if a terminal event happens first. The event counting process for the event $j$ with censoring is $N_{j}^a(t) = I\{\tilde{T}_{j}^a \leq t, \tilde{C}^a \geq t\} = I\{T^a\leq t, \Delta^a=j\}$ and the at-risk process is $Y^a(t) = I\{\tilde{T}^a\geq t, \tilde{C}^a\geq t\} = I\{T^a\geq t\}$. We adopt the ignorability assumption which is commonly used in causal inference literature. Ignorability means that the treatment assignment is independent of the potential event time and censoring time given the baseline covariates. 
\begin{assumption}[Ignorability] \label{ass:ign}
$A \independent (\tilde{T}^a, \tilde\Delta^a, \tilde{C}^a) \mid X$, for $a=0,1$.
\end{assumption}
As for censoring, we assume that the censoring is completely non-informative. The potential censoring time should be (unconditionally) independent of the potential event time.
\begin{assumption}[Completely random censoring] \label{ass:ran}
$\tilde{C}^a \independent (\tilde{T}^a, \tilde\Delta^a)$, for $a=0,1$.
\end{assumption}
Completely random censoring is essential for nonparametric estimation. Assumption \ref{ass:ran} holds if the potential censoring time is independent of $(\tilde{T}^a, \tilde\Delta^a,X)$. In a more general case, the hazards of $\tilde{C}^a$ and $\tilde{T}_j^a$ may depend on $X$. Suppose that $X=(X_1,X_2)$, with $X_1 \independent X_2$. If the hazard of $\tilde{C}^a$ is determined by $X_1$ but not $X_2$, while the hazard of $\tilde{T}_j^a$ is determined by $X_2$ but not $X_1$, then Assumption \ref{ass:ran} holds. The backdoor paths between $\tilde{C}^a$ and $\tilde{T}_j^a$ should be blocked on the causal graph.

Define the propensity score $e(a;x) = P(A=a \mid X=x)$. We need positivity for the propensity score and censoring distribution.
\begin{assumption}[Positivity] \label{ass:pos}
$0<c<e(a;X)<1-c<1$ and $P(Y^a(t^*)=1\mid A=a)>c$ for $a=0,1$ and a constant $c>0$.
\end{assumption}
The first part of positivity says that the propensity score is bounded away from 0 and 1. The second part of positivity says that there are still individuals at risk at the end of study, which guarantees the hazard function can be well defined in $[0,t^*]$. The probability that censoring has not happened at time $t^*$ should be positive, so there are available data to estimate the counterfactual hazards.

In the realized trial, let $T$ be the observed event time, and $\Delta$ be the observed event indicator. We assume consistency to link the potential values with observed values.
\begin{assumption}[Consistency] \label{ass:con}
$T = T^A$, $\Delta = \Delta^A$.
\end{assumption}
The observed event counting process $N_{j}(t) = I\{T\leq t, \Delta=j\}$ for the event $j$ and the observed at-risk process $Y(t) = I\{T\geq t\}$.
Suppose we have a sample including $n$ independent individuals, randomly drawn from a super-population. When necessary, we use the subscript $i=1,\ldots,n$ to represent the individual index. The observed data can be written as either $\{(A_i,X_i,T_i,\Delta_i): i=1,\ldots,n\}$ or $\{(A_i,X_i,N_{ij}(s),Y_i(s), 0 \leq s \leq t^*): i=1,\ldots,n\}$.

The counterfactual hazard of event $j$ describes the instantaneous risk of event $j$ under the treatment condition $a$,
\begin{align*}
d\Lambda_{j}^a(t) &= P(t\leq T_{j}^a <t+dt \mid T^a\geq t) \\
&= P(d\tilde{N}_{j}^a(t)=1 \mid Y^a(t)=1), \ 0\leq t<t^*,
\end{align*}
with $dt \to 0$. By transforming the hazards, the counterfactual cumulative incidence function (CIF) of event $j$ under the treatment condition $a$ is given by ($a=0,1$, $j=1,\ldots,J$)
\begin{equation}
F_j^a(t) = P(\tilde{T}^a \leq t, \tilde\Delta^a=j) = \int_0^t \exp\left\{-\sum_{k=1}^{J}\Lambda_k^a(s)\right\} d\Lambda_j^a(s), \ 0\leq t<t^*. \label{cif}
\end{equation}
We see that the hazards play the key role for time-to-event data analysis in that the CIF of any event can be derived from the hazards. The task is to identify and estimate the counterfactual hazard of each event under the treatment condition $a \in \{0,1\}$. A natural idea is to use inverse probability weighting (IPW). By weighting the individuals in the sample, a pseudo-population with baseline covariates distributed identically as the overall population is generated. Since the counterfactual hazard function is a conditional probability, IPW should be performed separately for incidence and at-risk probability. Denote $$w(a;A,X) = \frac{I\{A=a\}}{e(a;X)},$$ which is the inverse of the treatment propensity score multiplied by the associated treatment indicator. We have the following identifiability result.
\begin{theorem} \label{thm1}
Under Assumptions \ref{ass:ign}--\ref{ass:con}, the counterfactual cumulative cause-specific hazard of event $j$ is identifiable ($a=0,1$, $j=1,\ldots,J$),
\begin{align*}
\Lambda_j^a(t) = \int_0^t \frac{E\{w(a;A,X)dN_{j}(s)\}}{E\{w(a;A,X)Y(s)\}}.
\end{align*}
Thus, the counterfactual cumulative incidence function of event $j$ is identifiable.
\end{theorem}

Inspired by this theorem, if the propensity score is known, an estimator for the counterfactual cumulative cause-specific hazard by IPW is given by
\begin{equation}
\widetilde\Lambda_j^a(t) = \int_0^t \frac{\sum_{i=1}^{n}w(a;A_i,X_i)dN_{ij}(s)}{\sum_{i=1}^{n}w(a;A_i,X_i)Y_{i}(s)}.
\end{equation}
Otherwise, if the propensity score is unknown, we plug the estimated propensity score in $w(a;A_i,X_i)$ to obtain an empirical version $\widehat{w}(a;A_i,X_i)$. The resulting estimator for the counterfactual cumulative cause-specific hazard is given by
\begin{equation}
\widehat\Lambda_j^a(t) = \int_0^t \frac{\sum_{i=1}^{n}\widehat{w}(a;A_i,X_i)dN_{ij}(s)}{\sum_{i=1}^{n}\widehat{w}(a;A_i,X_i)Y_{i}(s)}.
\end{equation}
The estimator for the counterfactual cumulative hazard is a step function with jumps at event times. Based on Equation \eqref{cif}, we obtain the following adjusted Nelson--Aalen estimator for the counterfactual cumulative incidence function of event $j$ when the propensity score is estimated,
\begin{equation}
\widehat{F}_j^a(t) = \int_0^t \exp\left\{-\sum_{k=1}^{J} \widehat\Lambda_k^a(s)\right\} d\widehat\Lambda_j^a(s), \ 0 \leq t < t^*.
\end{equation}

\section{Asymptotic properties and inference} \label{sec3}

In this section, we derive the asymptotic variance of the adjusted Nelson--Aalen estimator. Let 
\begin{align*}
\Psi_{1j}(t;a) &= P(T^a\leq t, \Delta^a=j) = E\{w(a;A,X)N_{j}(t)\}, \\
\Psi_{2}(t;a) &= P(T^a \geq t) = E\{w(a;A,X)Y(t)\},
\end{align*}
then it follows from Theorem \ref{thm1} that $\Lambda_j^a(t)=\int_0^t d\Psi_{1j}(s;a)/\Psi_2(s;a)$. Let 
\begin{align*}
\psi_{1j}(t;a) &= w(a;A,X) N_j(t), \ \psi_2(t;a) = w(a;A,X) Y(t).
\end{align*}
We use $\mathbb{P}(\cdot)$ to denote the measure concerning the true data generating process and $\mathbb{P}_n(\cdot)$ to denote the empirical measure on the sample, so $\Psi_{1j}(t;a) = \mathbb{P}\{\psi_{1j}(t;a)\}$ and $\Psi_{2}(t;a) = \mathbb{P}\{\psi_{2}(t;a)\}$.

\subsection{When the propensity score is known}

We first assume that the propensity score is known. For example, individuals are assigned to treatment arms with a known probability in stratified randomized controlled trials. The oracle IPW estimator for the counterfactual cumulative cause-specific hazard of event $j$ is $\widetilde\Lambda_j^a(t) = \int_0^t \mathbb{P}_n\{d\psi_{1j}(s;a)\}/\mathbb{P}_n\{\psi_2(s;a)\}$. Let 
\[
M_j(t;a) = \psi_{1j}(t;a) - \int_0^t\psi_2(s;a)d\Lambda_j^a(s)
\]
and $\overline{M}_j(t;a) = \mathbb{P}_n\{M_j(t;a)\}$.
\begin{lemma} \label{lem1}
$\overline{M}_j(t;a)$ is a martingale with respect to the filter $\mathcal{F}_j(t;a) = \{w(a;A_i,X_i),Y_i(s): s\leq t, i=1,\ldots,n\}$.
\end{lemma}

Using the martingale theory, we can show that this oracle IPW estimator is unbiased and derive its variance.

\begin{theorem} \label{thm2}
Under Assumptions \ref{ass:ign}--\ref{ass:con} and that the propensity score is known, 
\[
E\{\widetilde\Lambda_j^a(t)\} = \Lambda_j^a(t), \ \var\{\widetilde\Lambda_j^a(t)\} = \frac{1}{n} E \left\{\int_0^t \frac{\mathbb{P}_n\{w(a;A,X)\psi_2(s;a)\}d\Lambda_j^a(s)}{[\mathbb{P}_n\{\psi_2(s;a)\}]^2} \right\}.
\]
The influence function (IF) of $\widetilde{\Lambda}_j^a(t)$ is
\[
\IF\{\widetilde\Lambda_j^a(t)\} = \int_0^t \frac{1}{\Psi_2(s;a)}dM_j(s;a),
\]
and thus
\[
\sqrt{n}\{\widetilde\Lambda_j^a(t)-\Lambda_j^a(t)\} \xrightarrow{d} N\left(0, E[\IF\{\widetilde\Lambda_j^a(t)\}^2]\right).
\]
\end{theorem}
The finite-sample variance of $\widetilde\Lambda_j^a(t)$ can be unbiasedly estimated by the empirical counterpart of the variance formula. The asymptotic variance can be consistently estimated by the variance of the influence function evaluated using observed data.

\subsection{When the propensity score is estimated}

Suppose that the propensity score $e(a;x)$ belongs to a model indexed by $\theta$, that is, $\{e(a;x;\theta),\theta \in \Theta\}$, where $\Theta$ is the parameter space. Furthermore, we assume $\{e(a;x;\theta),\theta \in \Theta\}$ is a Donsker class and the estimate of $\theta$ is regular and asymptotically linear (RAL), with 
\[
\widehat\theta - \theta = \mathbb{P}_n\phi + o_p(n^{-1/2}) = \frac{1}{n} \sum_{i=1}^{n} \phi(A_i,X_i;\theta) + o_p(n^{-1/2}),
\]
where $\phi$ is the influence function of $\widehat\theta$. This is to say that the model for the propensity score should not be too complex \citep{vaart2023empirical}. For example, the propensity score can be fitted by logistic model $e(1;x;\theta) = \{1+\exp(-x'\theta)\}^{-1}$. If $\theta$ is estimated by maximum likelihood, then $\phi = [\mathbb{P}\{Xe(1;X;\theta)e(0;X;\theta)X'\}]^{-1}\{A-e(1;X;\theta)\}X$. Plugging in the estimated propensity score $\widehat{e}(a;x)=e(a;x;\widehat\theta)$, let
\[
\widehat\psi_{1j}(t;a) = \frac{I\{A=a\}}{e(a;X;\widehat\theta)}N_j(t), \ \widehat\psi_2(t;a) = \frac{I\{A=a\}}{e(a;X;\widehat\theta)}Y(t),
\]
so the IPW estimator for the counterfactual cumulative cause-specific hazard of event $j$ is $\widehat\Lambda_j^a(t) = \int_0^t \mathbb{P}_n\{d\widehat\psi_{1j}(s)\}/\mathbb{P}_n\{\widehat\psi_2(s)\}$.

\begin{theorem} \label{thm3}
Under Assumptions \ref{ass:ign}--\ref{ass:con} and that $\widehat\theta$ in the propensity score model is RAL, the influence function (IF) of $\widehat\Lambda_j^a(t)$ is
\[
\IF\{\widehat\Lambda_j^a(t)\} = \int_0^t \frac{1}{\Psi_2(s;a)}\left[dM_j(s;a)-\mathbb{P}\left\{dM_j(s;a)\frac{\dot{e}(a;X;\theta)}{e(a;X;\theta)^2}\right\}\phi\right],
\]
and thus
\[
\sqrt{n}\{\widehat\Lambda_j^a(t)-\Lambda_j^a(t)\} \xrightarrow{d} N\left(0, E[\IF\{\widehat\Lambda_j^a(t)\}^2]\right).
\]
\end{theorem}

The proofs of the lemma and theorems are given in Supplementary Material A. 
Compared with the influence function of the oracle estimator, there is an augmented term in the influence function of $\widehat\Lambda_j^a(t)$, 
\[
\nu_j(t;a) = -\int_0^t \frac{1}{\Psi_2(s;a)}\mathbb{P}\left\{dM_j(s;a)\frac{\dot{e}(a;X;\theta)}{e(a;X;\theta)^2}\right\}\phi,
\]
where $\dot{e}(a;X;\theta)$ is the derivative of $e(a;X;\theta)$ with respect to $\theta$.
The expectation of this augmented term is not zero because $\{X_i: i=1,\ldots,n\}$ is not in the filter $\mathcal{F}_j(t;a)$. But by noticing that $\mathbb{P}_n\{M_j(t;a)\}$ is a martingale with mean zero, the expectation of this augmented term $\nu_j(t;a)$ is generally small. The variance of $\widehat\Lambda_j^a(t)$ can be consistently estimated by plugging the estimates into the influence function,
\begin{equation}
\widehat\sigma_{j,n}^{a,2}(t) = \frac{1}{n^2}\sum_{i=1}^{n}\left[\int_0^t \frac{1}{\mathbb{P}_n\{\widehat\psi_{2}(s;a)\}}\left[d\widehat{M}_{ij}(s;a)-\mathbb{P}_n\left\{d\widehat{M}_{j}(s;a)\frac{\dot{e}(a;X;\widehat\theta)}{e(a;X;\widehat\theta)^2}\right\}\phi(A_i,X_i;\widehat\theta)\right]\right]^2, \label{sigma_un}
\end{equation}
where $\widehat{M}_{ij}(t;a) = \widehat\psi_{1ij}(t;a) - \int_0^t\widehat\psi_{2i}(s;a) d\widehat\Lambda_j^a(s)$ is the martingale with estimates plugged in.
As a comparison, the variance estimate by ignoring the uncertainty of the estimated propensity score is as follows,
\begin{equation}
\widetilde\sigma_{j,n}^{a,2}(t) = \frac{1}{n^2}\sum_{i=1}^{n}\left[\int_0^t \frac{1}{\mathbb{P}_n\{\widehat\psi_{2}(s;a)\}}d\widehat{M}_{ij}(s;a)\right]^2. \label{sigma_kn}
\end{equation}

On the cumulative incidence scale, the influence function of $\widehat{F}_j^a(t)$ is derived by the functional delta method,
\[
\IF\{\widehat{F}_j^a(t)\} = \int_0^t \exp\left\{-\sum_{k=1}^{J}\Lambda_k^a(s)\right\} d\IF\{\widehat\Lambda_j^a(s)\} - \int_0^t \sum_{k=1}^{J}\IF\{\widehat\Lambda_k^a(s)\}dF_j(s),
\]
and thus
\[
\sqrt{n}\{\widehat{F}_j^a(t)-F_j^a(t)\} \xrightarrow{d} N\left(0, E[\IF\{\widehat{F}_j^a(t)\}]\right).
\]
The asymptotic variance of $\widehat{F}_j^a(t)$ can be estimated by plug-in estimators. Sometimes researchers are also interested in the average treatment effect (ATE) $\tau_j(t) = F_j^1(t)-F_j^0(t)$, the difference in the counterfactual cumulative incidences of event $j$ under the treated and control. The average treatment effect can be estimated by plug-in estimators, and the asymptotic properties can be easily obtained due to the additivity of influence functions.

\section{Simulation studies} \label{sec4}

In this section, we conduct simulation studies to assess the finite-sample performance of the adjusted Nelson--Aalen estimator and compare confidence intervals based on Equations \eqref{sigma_un} and \eqref{sigma_kn} when the propensity score is estimated.

Suppose that there are three covariates $X = (X_1,X_2,X_3)$ following the multivariate standard normal distribution. The potential failure times of the first occurrence under the treated and under the control are assumed to follow the proportional hazards model,
\begin{align*}
d\Lambda^1(t;x) &= (t^2/5) \exp(x_2/5+x_3/5) dt, \\
d\Lambda^0(t;a) &= (t/3) \exp(x_2/3-x_3/3) dt,
\end{align*}
from which we generate $\tilde{T}^1$ and $\tilde{T}^0$.
We assume there are two competing events. The probabilities that the first type of event (which is the event of interest) occurs are
\begin{align*}
P(\tilde\Delta^1=1\mid \tilde{T}^1=t, X=x) &= \expit(-0.1+0.2x_1+0.2x_2+0.2x_3+0.03t), \\
P(\tilde\Delta^0=1\mid \tilde{T}^0=t, X=x) &= \expit(0.2x_1-0.1x_2+0.1x_3+0.05t)
\end{align*}
under the treated and control, respectively, where $\expit(x)=1/\{1+\exp(-x)\}$. The censoring time is generated from a uniform distribution in $[6,12]$. The propensity score is assumed to follow the logistic model,
\[
P(A=1 \mid X=x) = \expit(0.2+0.5x_1-0.5x_2).
\]

To examine the empirical bias, we compare three point estimates: (1) the oracle adjusted Nelson--Aalen (NA) estimate that uses the true propensity score, (2) the adjusted Nelson--Aalen estimate that uses the estimated propensity score, and (3) the weighted Aalen--Johansen (AJ) estimate \citep{aalen1978empirical} by imposing the inverse of propensity score as weights in the function \texttt{survfit} in library \texttt{survival} of R \citep{survival-package}. 
We consider five methods to obtain the standard error: (1) the oracle standard error that uses the true propensity score in Theorem \ref{thm2}, (2) the naive standard error $\widetilde\sigma_{j,n}^a(t)$ that does not consider the uncertainty of the estimated propensity score, (3) the standard error $\widehat\sigma_{j,n}^a(t)$ that corrects for the estimated propensity score in Theorem \ref{thm3}, (4) the nonparametric bootstrap standard error, and (5) the standard error of the weighted Aalen-Johansen estimate calculated by \texttt{survfit}.

Next, we consider five confidence intervals: (1) the oracle confidence interval using the oracle adjusted Nelson--Aalen estimate and the oracle standard error, (2) the naive confidence interval using the adjusted Nelson--Aalen estimate and the naive standard error, (3) the corrected confidence interval using the adjusted Nelson--Aalen estimate and the corrected standard error, (4) the bootstrap confidence interval using the adjusted Nelson--Aalen estimate and the bootstrap standard error, and (5) the confidence interval using the weighted Aalen--Johansen estimate and associated standard error in \texttt{survfit}. Nominal 95\% confidence intervals are constructed using the point estimate and standard error based on asymptotic normality.

\begin{figure}
\centering
\includegraphics[width=0.9\textwidth]{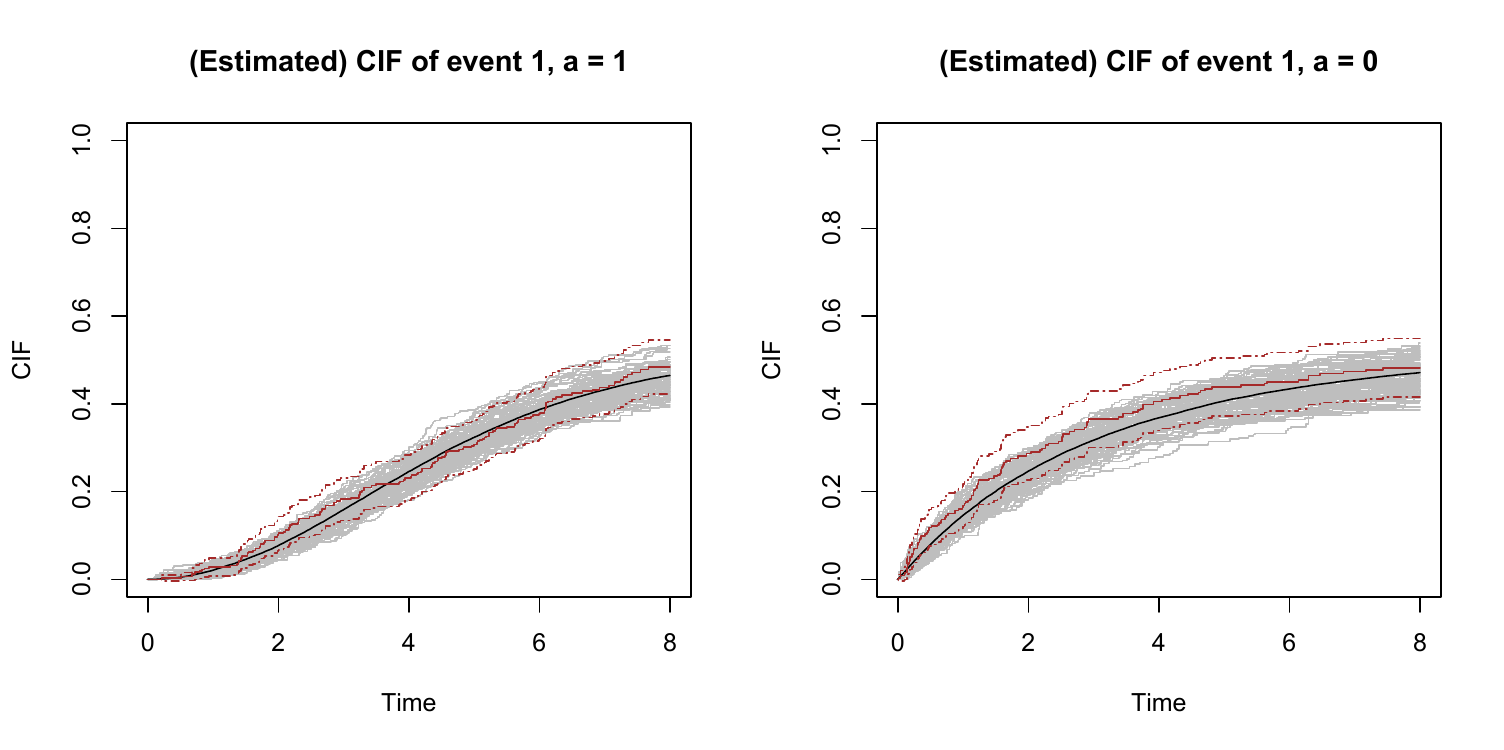}
\caption{The estimated counterfactual cumulative incidences by the adjusted Nelson--Aalen estimator. The black line is the true value, the grey lines are the estimates (we draw 100 lines). The solid brown line is a single estimate, and the dashed brown lines are the upper and lower bounds of the 95\% confidence interval associated with this estimate.} \label{fig:est}
\end{figure}

Figure \ref{fig:est} shows 100 adjusted Nelson--Aalen estimates of the counterfactual cumulative incidence $F_1^a(t)$, $a=1,0$ when the sample size $n=500$. The estimates distribute around the true value.
Table \ref{tab:F1_500} and Table \ref{tab:F0_500} show the empirical bias of the point estimates with standard deviation, mean standard error and coverage rate of the nominal 95\% confidence intervals when the sample size $n=500$. We see negligible bias in the adjusted Nelson--Aalen estimator. Even if we use the estimated propensity score, the standard deviation (and standard error) is similar to the oracle one, indicating that the efficiency loss due to estimating the propensity score is negligible. Considering the uncertainty of the estimated propensity score, the standard error based on the influence function in Theorem \ref{thm3} is slightly larger than the naive one, and hence the coverage rate based on the corrected standard error is larger. The weighted Aalen--Johansen estimator from the R function \texttt{survfit} gives a similar standard error to the naive adjusted Nelson--Aalen estimator, and these two types of estimators are actually asymptotically equivalent with slightly different finite-sample performances \citep{luo1993bias, colosimo2002empirical}. The most interesting finding is that omitting the variation of the estimated propensity score does not lead to much bias of the standard error. In Supplementary Material C, we provide more simulation results, including the setting where the sample size is $n=2000$ and a sensitivity analysis where the propensity score model is misspecified.

\begin{table}
\centering
\caption{Bias, standard deviation, mean standard error and confidence interval coverage rate of some estimators for the counterfactual cumulative incidence function $F_1^1(t)$ when the sample size $n=500$} \label{tab:F1_500}
\begin{tabular}{lcccccccc}
  \toprule
Time & 1 & 2 & 3 & 4 & 5 & 6 & 7 & 8 \\ 
  \midrule
  \multicolumn{9}{l}{Bias} \\
Oracle & -0.000 & -0.001 & -0.001 & -0.000 & -0.000 & -0.001 & -0.002 & -0.003 \\ 
Adjusted NA & -0.000 & -0.001 & -0.001 & -0.000 & -0.001 & -0.001 & -0.002 & -0.003 \\ 
Weighted AJ & -0.000 & -0.000 & 0.000 & 0.001 & 0.001 & 0.002 & 0.001 & 0.001 \\ 
\midrule
\multicolumn{9}{l}{Standard deviation (SD)} \\
Oracle & 0.009 & 0.017 & 0.024 & 0.028 & 0.029 & 0.029 & 0.030 & 0.032 \\ 
Adjusted NA & 0.009 & 0.017 & 0.024 & 0.028 & 0.029 & 0.030 & 0.030 & 0.032 \\ 
Weighted AJ & 0.009 & 0.018 & 0.024 & 0.028 & 0.029 & 0.030 & 0.031 & 0.032 \\ 
\midrule
\multicolumn{9}{l}{Standard error (SE)} \\
Oracle & 0.009 & 0.017 & 0.023 & 0.027 & 0.030 & 0.031 & 0.031 & 0.032 \\ 
Naive & 0.009 & 0.017 & 0.023 & 0.028 & 0.030 & 0.031 & 0.032 & 0.032 \\ 
Corrected & 0.009 & 0.019 & 0.027 & 0.031 & 0.033 & 0.035 & 0.035 & 0.035 \\ 
Bootstrap & 0.009 & 0.017 & 0.023 & 0.027 & 0.029 & 0.030 & 0.031 & 0.031 \\ 
Weighted AJ & 0.009 & 0.017 & 0.024 & 0.028 & 0.030 & 0.031 & 0.032 & 0.033 \\ 
\midrule
\multicolumn{9}{l}{Coverage rate} \\
Oracle & 0.886 & 0.934 & 0.950 & 0.941 & 0.955 & 0.955 & 0.957 & 0.955 \\ 
Naive & 0.887 & 0.935 & 0.946 & 0.946 & 0.957 & 0.957 & 0.966 & 0.952 \\ 
Corrected & 0.889 & 0.942 & 0.959 & 0.957 & 0.967 & 0.969 & 0.973 & 0.968 \\ 
Bootstrap & 0.885 & 0.933 & 0.939 & 0.936 & 0.952 & 0.953 & 0.958 & 0.948 \\ 
Weighted AJ & 0.887 & 0.938 & 0.950 & 0.946 & 0.958 & 0.963 & 0.965 & 0.958 \\ 
   \bottomrule
\end{tabular}
\end{table}

\begin{table}
\centering
\caption{Bias, standard deviation, mean standard error and confidence interval coverage rate of some estimators for the counterfactual cumulative incidence function $F_1^0(t)$ when the sample size $n=500$} \label{tab:F0_500}
\begin{tabular}{lcccccccc}
  \toprule
Time & 1 & 2 & 3 & 4 & 5 & 6 & 7 & 8 \\ 
  \midrule
  \multicolumn{9}{l}{Bias} \\
Oracle & -0.001 & -0.002 & -0.003 & -0.003 & -0.003 & -0.004 & -0.005 & -0.005 \\ 
Adjusted NA & -0.001 & -0.002 & -0.003 & -0.003 & -0.003 & -0.004 & -0.005 & -0.005 \\ 
Weighted AJ & 0.000 & -0.000 & -0.000 & 0.000 & 0.000 & 0.000 & -0.000 & -0.000 \\  
\midrule
  \multicolumn{9}{l}{Standard deviation (SD)} \\
Oracle & 0.025 & 0.031 & 0.034 & 0.034 & 0.035 & 0.035 & 0.036 & 0.037 \\ 
Adjusted NA & 0.025 & 0.031 & 0.033 & 0.034 & 0.035 & 0.036 & 0.037 & 0.037 \\ 
Weighted AJ & 0.025 & 0.031 & 0.034 & 0.034 & 0.035 & 0.036 & 0.037 & 0.037 \\ 
\midrule
  \multicolumn{9}{l}{Standard error (SE)} \\
Oracle & 0.024 & 0.030 & 0.033 & 0.034 & 0.034 & 0.035 & 0.035 & 0.035 \\ 
Naive & 0.025 & 0.030 & 0.033 & 0.034 & 0.035 & 0.035 & 0.035 & 0.035 \\ 
Corrected & 0.025 & 0.031 & 0.033 & 0.035 & 0.035 & 0.036 & 0.036 & 0.036 \\ 
Bootstrap & 0.024 & 0.030 & 0.032 & 0.033 & 0.034 & 0.034 & 0.034 & 0.035 \\ 
Weighted AJ & 0.025 & 0.031 & 0.033 & 0.035 & 0.035 & 0.036 & 0.036 & 0.036 \\ 
\midrule
  \multicolumn{9}{l}{Coverage rate} \\
Oracle & 0.937 & 0.942 & 0.931 & 0.949 & 0.937 & 0.942 & 0.938 & 0.941 \\ 
Naive & 0.932 & 0.946 & 0.932 & 0.947 & 0.941 & 0.943 & 0.934 & 0.941 \\ 
Corrected & 0.941 & 0.951 & 0.935 & 0.947 & 0.943 & 0.950 & 0.942 & 0.945 \\ 
Bootstrap & 0.934 & 0.942 & 0.927 & 0.940 & 0.940 & 0.939 & 0.934 & 0.935 \\ 
Weighted AJ & 0.939 & 0.951 & 0.940 & 0.948 & 0.947 & 0.951 & 0.941 & 0.944 \\ 
   \bottomrule
\end{tabular}
\end{table}

\section{Application to allogeneic stem cell transplantation data} \label{sec5}

Allogeneic stem cell transplantation is a widely applied therapy to treat acute lymphoblastic leukemia (ALL), including two sorts of transplant modalities: human leukocyte antigens matched sibling donor transplantation (MSDT) and haploidentical stem cell transplantation from family (Haplo-SCT). MSDT has long been regarded as the first choice of transplantation because MSDT leads to lower transplant-related mortality, also known as non-relapse mortality (NRM) \citep{kanakry2016modern}. In recent years, some benefits of Haplo-SCT have been noticed that patients with positive pre-transplantation minimum residual disease (MRD) undergoing Haplo-SCT have better prognosis in relapse, and hence lower relapse-related mortality \citep{chang2020haploidentical}. It is interesting to investigate whether Haplo-SCT can be an alternative to MSDT, since the former is much more accessible. We adopt the relapse as the primary event and non-relapse mortality as the competing event.

A total of $n=303$ patients with positive MRD undergoing allogeneic stem cell transplantation at Peking University People's Hospital in China from 2009 to 2017 were included in our study \citep{ma2021an}. Among these patients, 65 received MSDT ($A=0$) and 238 received Haplo-SCT ($A=1$). There is no specific consideration to prefer Haplo-SCT over MSDT whenever MSDT is accessible \citep{chang2020haploidentical}, so we expect ignorability. Four baseline covariates are considered: age, sex (male or female), diagnosis (T-cell ALL [T-ALL] or B-cell ALL [B-ALL]) and complete remission status (after 1 cycle [CR1] or more than 1 cycle [CR$>$1]). As found in previous literature, these covariates are risk factors associated with relapse and mortality. The time origin is the time of receiving transplantation (either MSDT or Haplo-SCT). The mean follow-up time is 1336 days, and the maximum follow-up time is 4106 days. In the MSDT group, 47.7\% patients were observed to encounter relapse and 9.2\% NRM. In the Haplo-SCT group, 29.0\% patients were observed to encounter relapse and 11.8\% NRM. Summary statistics are presented in Table \ref{tab:sum}.

\begin{table}
\centering
\caption{Summary statistics in the stem cell transplantation data, stratified by the treatment groups. We list the mean and standard deviation (SD) for continuous variables, and list the count and proportion for binary variables} \label{tab:sum}
\begin{tabular}{lrrrr}
  \toprule
 & \multicolumn{2}{c}{Haplo-SCT ($A=1$)} & \multicolumn{2}{c}{MSDT ($A=0$)} \\ 
 \cmidrule(lr){2-3} \cmidrule(lr){4-5}
 & Mean/Count & SD/Proportion & Mean/Count & SD/Proportion \\
  \midrule
  \multicolumn{5}{l}{Baseline covariates} \\
  Age & 26.7 & (12.2) & 35.0 & (13.1) \\ 
  Sex: Male & 89 & 37.4\% & 27 & 41.5\% \\ 
  Complete Remission: CR1 & 54 & 22.7\% & 10 & 15.4\% \\ 
  Diagnosis: T-ALL & 38 & 16.0\% & 3 & 4.6\% \\ 
  \midrule
  \multicolumn{5}{l}{Observed events of first occurrence} \\
  Relapse & 69 & 29.0\% & 31 & 47.7\% \\ 
  Time to relapse (days) & 371.7 & (406.2) & 420.8 & (369.8) \\ 
  NRM & 28 & 11.8\% & 6 & 9.2\% \\ 
  Time to NRM (days) & 192.2 & (192.0) & 207.5 & (95.2) \\ 
   \bottomrule
\end{tabular}
\end{table}

\begin{figure}
\centering
\includegraphics[width=0.9\textwidth]{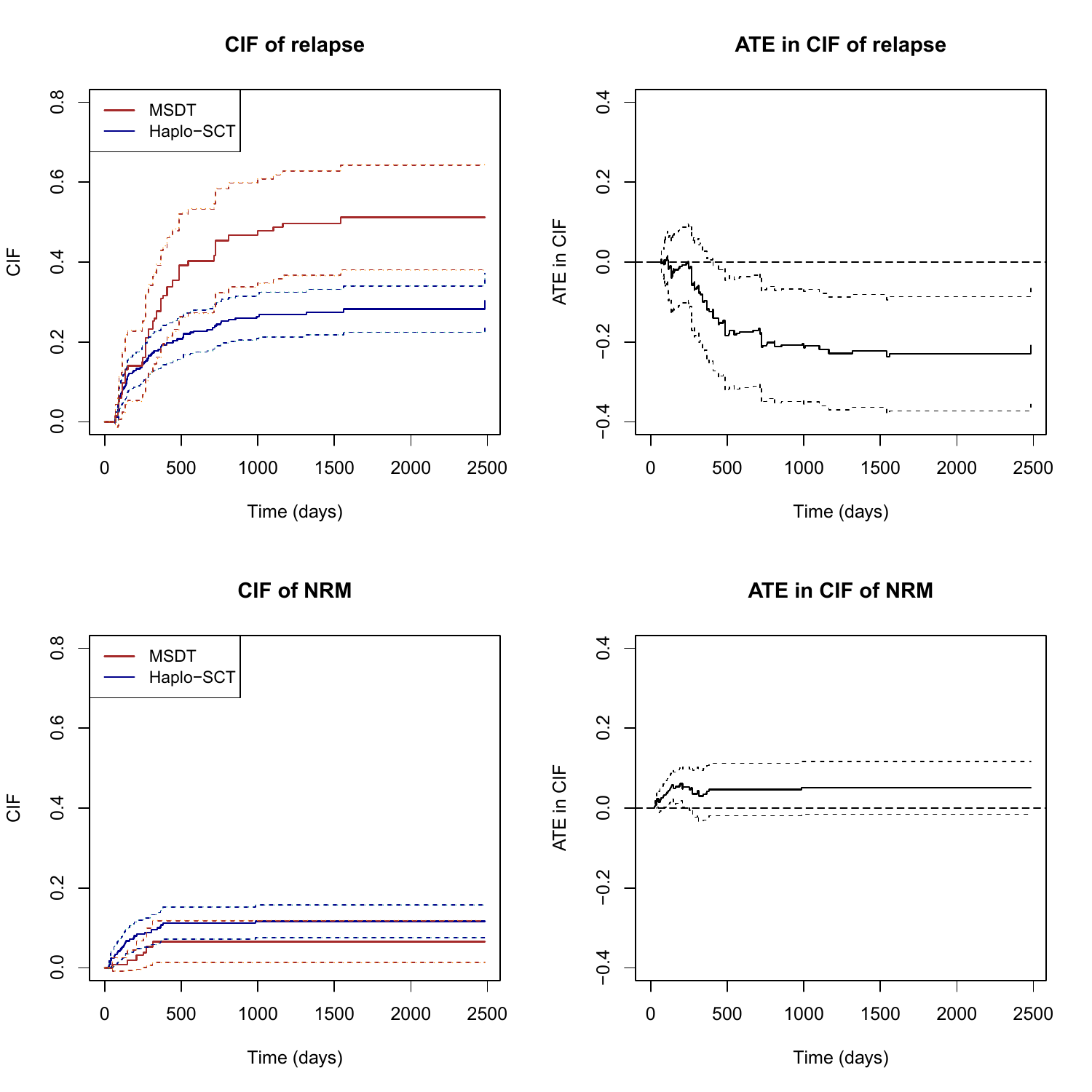}
\caption{Left: The estimated counterfactual cumulative incidence functions of relapse and non-relapse mortality (NRM). The brown line is MSDT and the blue line is Haplo-SCT. The brown and blue dashed lines are the 95\% confidence intervals by correcting for the estimated propensity score; the orange and cyan lines are the 95\% confidence intervals ignoring the uncertainty of the estimated propensity score. Right: the average treatment effect (ATE) on the incidence scale. Negative values indicate that Haplo-SCT leads to a lower relapse (NRM) rate than MSDT.} \label{fig:cif}
\end{figure}

The propensity score is fitted by logistic regression. The upper-left panel in Figure \ref{fig:cif} shows the estimated counterfactual cumulative incidence function of relapse. The relapse rate is higher in the MSDT group than in the Haplo-SCT group. The 95\% confidence intervals are displayed in dashed lines. We plot the confidence interval that ignores the uncertainty of the estimated propensity score and the confidence interval that corrects for the estimated propensity score. The numerical values of these two types of confidence intervals are very similar, and we cannot visually tell a difference between these two types of confidence intervals.
The upper-right panel of Figure \ref{fig:cif} presents the average treatment effect (ATE) of Haplo-SCT on the relapse rate compared to MSDT, defined as the difference in counterfactual cumulative incidences. The 95\% confidence interval calculated based on the influence functions of the estimated CIFs is displayed in dashed lines. Negative values of the ATE indicate that Haplo-SCT leads to a lower relapse rate than MSDT. The bottom-left panel shows the estimated cumulative incidence function of NRM and the bottom-right panel shows the estimated average treatment effect. The NRM rate is slightly higher in the Haplo-SCT group than in the MSDT group.
We list the estimated treatment effects and 95\% confidence intervals at some time points in Table \ref{tab:ate}.

\begin{table}
\centering
\caption{Average treatment effects (ATEs) on the counterfactual cumulative incidence functions of relapse and NRM with 95\% confidence intervals (CIs) at some time points} \label{tab:ate}
\begin{tabular}{lccccccc}
\toprule
Years after & \multicolumn{3}{c}{Relapse} & \multicolumn{3}{c}{NRM} \\
\cmidrule(lr){2-4} \cmidrule(lr){5-7}
transplantation & ATE & 95\% CI & $P$-value & ATE & 95\% CI & $P$-value \\
\midrule
0.5 & $-0.015$ & $(-0.109, 0.079)$ & 0.752 & $0.054$ & $(0.012, 0.095)$ & 0.012 \\
1.0 & $-0.096$ & $(-0.220, 0.027)$ & 0.126 & $0.038$ & $(-0.026, 0.103)$ & 0.244 \\
2.0 & $-0.209$ & $(-0.349, -0.069)$ & 0.003 & $0.046$ & $(-0.019, 0.112)$ & 0.164 \\
3.0 & $-0.209$ & $(-0.350, -0.069)$ & 0.004 & $0.051$ & $(-0.015, 0.117)$ & 0.131 \\
5.0 & $-0.230$ & $(-0.372, -0.087)$ & 0.002 & $0.051$ & $(-0.015, 0.117)$ & 0.131 \\
\bottomrule
\end{tabular}
\end{table}

In summary, we find that Haplo-SCT significantly reduces the risk of relapse compared to MSDT. Haplo-SCT can be used as the better transplant modality for MRD positive patients. Since Haplo-SCT is more accessible than MSDT, it is promising that Haplo-SCT be considered as the first choice of allogeneic stem cell transplantation as long as paying more care to prevent transplant-related mortality.

\section{Discussion} \label{sec6}

Inverse probability weighting has been well applied in causal inference due to its simplicity and interpretability. However, the estimand (cumulative incidence function) is time-varying for time-to-event outcomes. We may employ IPW in two ways: the first is to directly estimate the incidence by weighting the observed event counting process by the inverse of propensity score and uncensored probability, and the second is to estimate the population-level hazard by weighting both the event counting process and at-risk process. In this article, we follow the second approach. 

There are a few advantages of estimating the cumulative incidence through transforming hazards over directly targeting the cumulative incidence. First, only one model (the treatment propensity score) is required. We do not need to estimate the censoring probability as long as the censoring is completely non-informative or estimate the conditional hazards with complicated post-treatment (time-varying) covariates. Second, the additional variance resulted by the uncertainty of the estimated propensity score is negligible since the extra term in the influence function of $\widehat\Lambda_j^a(t)$ is a weighted martingale, although the expectation may not be exactly zero.

To conclude, we point out five possible extensions for the proposed method.
First, there are alternatives on the estimation of the propensity score. Different parametric models like logistic regression and probit regression can be used. Since IPW essentially uses the balancing property of the propensity score, we can adapt the loss functions to obtain the covariates balancing propensity score \citep{imai2014covariate}. As long as the estimated propensity score has a known form of influence function, the asymptotic variance of the adjusted Nelson--Aalen estimator can be corrected.

Second, estimating the hazard is potentially helpful for mediation analysis. Within the interventional effects framework, suppose we can draw the event counting process from some reference distribution (such as according to the counterfactual cause-specific hazard under control). Under some additional assumptions, we can identify and estimate the counterfactual cumulative incidence of each event under such an intervention. This allows one to study the direct and indirect effects on a single event. Testing the treatment effect can be reduced to an implication of testing the counterfactual cause-specific hazards, which can be achieved by logrank tests. However, the uncertainty of the estimated score should be accounted for in the tests.

Third, the competing risks framework can be extended to multi-state models. Typical assumptions in multi-state models are Markovness or semi-Markovness \citep{commenges2007choice, asanjarani2022estimation}. The former says that the transition hazard from one state to another only relies on the time since the origin rather than the history. In contrast, the latter says that the transition hazard only depends on how long it has passed since reaching the last state. With completely non-informative censoring, the transition hazards can be consistently estimated by IPW, and the cumulative incidence of any state can then be estimated after some standard derivation. The asymptotic properties of the estimated incidence can be established using the functional delta method. 

Fourth, the assumptions in our framework may be relaxed. If the censoring is not completely random but depends on observed covariates, the counterfactual hazard is still identifiable but more complicated. In addition to the propensity score, the censoring probability should also be adjusted to reflect the overall population. Informative censoring induces a biased selection issue, where the at-risk individuals may have different underlying features in the real world (subject to censoring) and counterfactual world (not subject to censoring). Therefore, the counting processes should be additionally weighted by the conditional uncensored probability, which introduces another source of variance. Suppose the estimated conditional uncensored probability is RAL (for example, by Cox regression). In that case, we can imitate the same strategy considered in this article to derive the asymptotic variance of the estimated counterfactual CIF.

Fifth, the efficiency can be improved by incorporating models for the failure times. We can derive the efficient influence function (EIF) for the counterfactual CIF under conditionally random censoring (which is weaker than completely random censoring), which involves the propensity score, censoring probability and cause-specific hazards for all events \citep{zhang2012contrasting, martinussen2023estimation, rytgaard2024targeted}. The explicit form of the EIF is given in the Supplementary Material B. This estimator based on EIF does not have the adjusted Nelson--Aalen form. In practice, the propensity score is fitted at baseline whereas the censoring probability and cause-specific hazards are fitted using post-treatment data. Modeling the univariate propensity score is computationally easier by regression and provides desirable asymptotic properties. It is challenging to correctly specify the hazards if the dependence of events is complex or if there are time-varying covariates. Misspecification of working models may lead to bias and inconsistent variance estimation.

\section*{Acknowledgments}
We thank Dr. Wang Miao (Peking University) for comments. We also thank Dr. Yingjun Chang, Dr. Leqing Cao and Dr. Yuewen Wang (Peking University People's Hospital) for cleaning the data.

\section*{Conflict of interest}
The authors declare no conflict of interest.

\section*{Supplementary material}
Supplementary material includes (A) proofs of theoretical results, (B) efficient influence function of the counterfactual cumulative incidence and (C) additional simulation results.

\includepdf[pages=1-13]{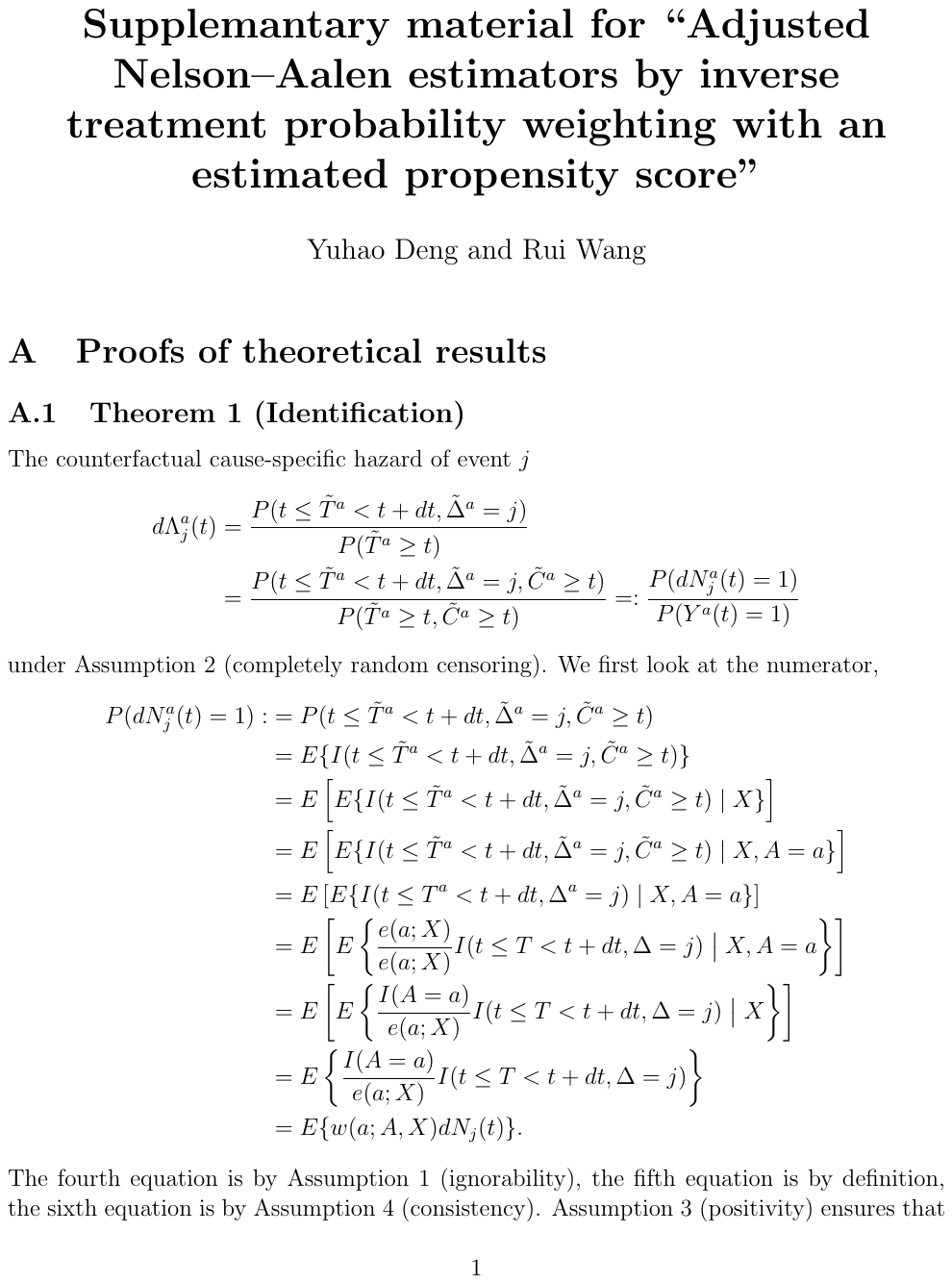}

\end{document}


\maketitle

\renewcommand{\thesection}{\Alph{section}}
\renewcommand{\theequation}{S\arabic{equation}}
\renewcommand{\thetable}{S\arabic{table}}
\renewcommand{\thefigure}{S\arabic{figure}}
\allowdisplaybreaks[3]

\section{Proofs of theoretical results}

\subsection{Theorem 1 (Identification)}

The counterfactual cause-specific hazard of event $j$
\begin{align*}
d\Lambda_j^a(t) &= \frac{P(t\le\tilde{T}^a<t+dt, \tilde\Delta^a=j)}{P(\tilde{T}^a \ge t)} \\
&= \frac{P(t\le\tilde{T}^a<t+dt, \tilde\Delta^a=j, \tilde{C}^a\ge t)}{P(\tilde{T}^a \ge t, \tilde{C}^a \ge t)} =: \frac{P(dN_j^a(t)=1)}{P(Y^a(t)=1)}
\end{align*}
under Assumption 2 (completely random censoring). We first look at the numerator,
\begin{align*}
P(dN_j^a(t)=1) :&= P(t\le\tilde{T}^a<t+dt, \tilde\Delta^a=j, \tilde{C}^a \ge t) \\
&= E\{I(t\leq \tilde{T}^a < t+dt, \tilde\Delta^a=j, \tilde{C}^a \ge t)\} \\
&= E\left[E\{I(t\leq \tilde{T}^a < t+dt, \tilde\Delta^a=j, \tilde{C}^a \ge t) \mid X \}\right] \\
&= E\left[E\{I(t\leq \tilde{T}^a < t+dt, \tilde\Delta^a=j, \tilde{C}^a \ge t) \mid X, A=a\}\right] \\
&= E\left[E\{I(t\leq T^a < t+dt, \Delta^a=j) \mid X, A=a\}\right] \\
&= E\left[E\left\{\frac{e(a;X)}{e(a;X)}I(t\leq T < t+dt, \Delta=j) \bigm| X, A=a\right\}\right] \\
&= E\left[E\left\{\frac{I(A=a)}{e(a;X)}I(t\leq T < t+dt, \Delta=j) \bigm| X\right\}\right] \\
&= E\left\{\frac{I(A=a)}{e(a;X)}I(t\leq T < t+dt, \Delta=j)\right\} \\
&= E\{w(a;A,X)dN_j(t)\}.
\end{align*}
The fourth equation is by Assumption 1 (ignorability), the fifth equation is by definition, the sixth equation is by Assumption 4 (consistency). Assumption 3 (positivity) ensures that the weight is well defined.
Similarly, for the denominator,
\begin{align*}
P(Y^a(t)=1) :&= P(\tilde{T}^a \geq t, \tilde{C}^a \geq t) \\
&= E\{I(\tilde{T}^a \geq t, \tilde{C}^a \geq t)\} \\
&= E\left[E\{I(\tilde{T}^a \geq t, \tilde{C}^a \geq t) \mid X\}\right] \\
&= E\left[E\{I(\tilde{T}^a \geq t, \tilde{C}^a \geq t) \mid X, A=a\}\right] \\
&= E\left[E\{I(T^a \geq t) \mid X, A=a\}\right] \\
&= E\left[E\left\{\frac{e(a;X)}{e(a;X)}I(T \geq t) \mid X, A=a\right\}\right] \\
&= E\left[E\left\{\frac{I(A=a)}{e(a;X)}I(T \geq t) \mid X\right\}\right] \\
&= E\left[\frac{I(A=a)}{e(a;X)}I(T \geq t) \right] \\
&= E\{w(a;A,X)Y(t)\}.
\end{align*}
Assumption 3 (positivity) ensures that $P(Y^a(t)=1) > 0$.
Therefore,
\[
d\Lambda_j^a(t) = \frac{E\{w(a;A,X)dN_j(t)\}}{E\{w(a;A,X)Y(t)\}}.
\]

\subsection{Lemma 1 (Martingale)}

Let $\mathcal{X}$ be the support of $X$. Without loss of generality, we assume that $X$ is discrete. The key to prove the martingale is to relate the population-level hazard $d\Lambda_j^a(t)$ with the conditional hazard
\begin{align*}
d\Lambda_j^a(t;x) &= P(dN_j^a(t)=1 \mid Y^a(t)=1, X=x) \\
&= P(dN_j(t)=1 \mid Y(t)=1, X=x, A=a).
\end{align*}
By iterated expectation, we examine $\overline{M}_j(t;a)$ in a refined filter,
\begin{align*}
&\quad~ E\{d\overline{M}_j(t;a) \mid \mathcal{F}_j(t;a)\} \\
&= E[\mathbb{P}_nd\{\psi_{1j}(t;a)-\psi_{2}(t;a)d\Lambda_j^a(t)\} \mid \mathcal{F}_j(t;a)] \\
&= E\left[E\{\mathbb{P}_n\{d\psi_{1j}(t;a)-\psi_{2}(t;a)d\Lambda_j^a(t)\} \mid \mathcal{F}_j(t;a), X_1, \ldots, X_n \} \mid \mathcal{F}_j(t;a)\right] \\
&= E\left[E\left\{ \frac{1}{n}\sum_{i=1}^{n}w(a;A_i,X_i)\{dN_{ij}(t)-Y_i(t)d\Lambda_j^a(t)\} \mid \mathcal{F}_j(t;a), X_1, \ldots, X_n \right\} \bigm| \mathcal{F}_j(t;a)\right] \\
&= E\left[\frac{1}{n}\sum_{i=1}^{n}w(a;A_i,X_i)\{Y_i(t)d\Lambda_j^a(t;X_i)-Y_i(t)d\Lambda_j^a(t)\} \bigm| \mathcal{F}_j(t;a) \right] \\
&= \frac{1}{n}\sum_{i=1}^{n}w(a;A_i,X_i)Y_i(t)E\{d\Lambda_j^a(t;X_i) \mid \mathcal{F}_j(t;a)\} - \frac{1}{n}\sum_{i=1}^{n}w(a;A_i,X_i)Y_i(t)d\Lambda_j^a(t) \\
&= \frac{1}{n}\sum_{i=1}^{n}w(a;A_i,X_i)Y_i(t) \int_{\mathcal{X}} P(X_i=x\mid Y_i(t), w(a;A_i,X_i))d\Lambda_j^a(t;x)dx \\
&\quad - \frac{1}{n}\sum_{i=1}^{n}w(a;A_i,X_i)Y_i(t)d\Lambda_j^a(t).
\end{align*}
We only need to care about the case $Y_i(t)=1$.
\begin{align*}
&\quad~ E\{d\overline{M}_j(t;a) \mid \mathcal{F}_j(t;a)\} \\
&= \frac{1}{n}\sum_{i=1}^{n}w(a;A_i,X_i)Y_i(t) \int_{\mathcal{X}} P(X_i=x\mid Y_i(t)=1, w(a;A_i,X_i))d\Lambda_j^a(t;x)dx \\
&\qquad\qquad- \frac{1}{n}\sum_{i=1}^{n}w(a;A_i,X_i)Y_i(t)d\Lambda_j^a(t) \\
&= \frac{1}{n}\sum_{i=1}^{n}w(a;A_i,X_i)Y_i(t) \int_{\mathcal{X}} P(X_i=x\mid Y_i(t)=1, w(a;A_i,X_i)) \frac{P(dN_{ij}^a(t)=1 \mid X_i=x)}{P(Y_i^a(t)=1 \mid X_i=x)}dx \\
&\qquad\qquad - \frac{1}{n}\sum_{i=1}^{n}w(a;A_i,X_i)Y_i(t)d\Lambda_j^a(t).
\end{align*}
When $A_i=a$ under which $w(a;A_i,X_i)\neq0$, we have that $Y_i^a(t)=Y_i(t)$, $N_{ij}^a(t)=N_{ij}(t)$ and $w(a,A_i,X_i)=1/e(a;X_i)$ is a function of $X_i$, so
\begin{align*}
&~\quad P(X_i=x\mid Y_i(t)=1, w(a;A_i,X_i)) \\
&= \frac{P(Y_i(t)=1\mid X_i=x, w(a;A_i,X_i)) P(X_i=x, w(a;A_i,X_i))}{\int_{\mathcal{X}}P(Y_i(t)=1\mid X_i=x, w(a;A_i,X_i)) P(X_i=x, w(a;A_i,X_i))dx} \\
&= \frac{P(Y_i^a(t)=1\mid X_i=x, e(X_i)) P(X_i=x, e(X_i))}{\int_{\mathcal{X}}P(Y_i^a(t)=1\mid X_i=x, e(X_i)) P(X_i=x, e(X_i))dx} \\
&= \frac{P(Y_i^a(t)=1\mid X_i=x)P(X_i=x)}{\int_{\mathcal{X}}P(Y_i^a(t)=1\mid X_i=x)P(X_i=x)dx} \\
&= \frac{P(Y_i^a(t)=1\mid X_i=x)P(X_i=x)}{P(Y_i^a(t)=1)}.
\end{align*}
Therefore
\begin{align*}
&\quad~ E\{d\overline{M}_j(t;a) \mid \mathcal{F}_j(t;a)\} \\
&= \frac{1}{n}\sum_{i=1}^{n}w(a;A_i,X_i)Y_i(t) \int_{\mathcal{X}} \frac{P(dN_{ij}^a(t)=1 \mid X_i=x)P(X_i=x)}{P(Y_i^a(t)=1)}dx \\
&\quad - \frac{1}{n}\sum_{i=1}^{n}w(a;A_i,X_i)Y_i(t)d\Lambda^a_j(t) \\
&= \frac{1}{n}\sum_{i=1}^{n}w(a;A_i,X_i)Y_i(t) \frac{P(dN_{ij}^a(t)=1)}{P(Y_i^a(t)=1)} - \frac{1}{n}\sum_{i=1}^{n}w(a;A_i,X_i)Y_i(t)d\Lambda_j^a(t) \\
&= \frac{1}{n}\sum_{i=1}^{n}w_i(a;A_i,X_i)Y_i(t)d\Lambda^a_j(t) - \frac{1}{n}\sum_{i=1}^{n}w_i(a;A_i,X_i)Y_i(t)d\Lambda^a_j(t) \\
&= 0.
\end{align*}
This proves that $\overline{M}_j(t;a)$ is a martingale.

\subsection{Theorem 2 (IF with known propensity score)}

Note that $\mathbb{P}_n\psi_2(t;a)$ is a predictable process. To see the unbiasedness of $\widetilde\Lambda_j^a(t)$,
\begin{align*}
E\{\widetilde\Lambda_j^a(t)\} &= E\left\{\int_0^t\frac{\mathbb{P}_nd\psi_{1j}(s;a)}{\mathbb{P}_n\psi_2(s;a)}\right\} \\
&= E\left[E\left\{\int_0^t\frac{\mathbb{P}_nd\psi_{1j}(s;a)}{\mathbb{P}_n\psi_2(s;a)} \bigm| \mathcal{F}_j(s;a)\right\}\right] \\
&= E\left[\int_0^t\frac{E\{\mathbb{P}_nd\psi_{1j}(s;a)\mid\mathcal{F}_j(s;a)\}}{\mathbb{P}_n\psi_2(s;a)}\right] \\
&= E\left\{\int_0^t\frac{\mathbb{P}_n\psi_2(s;a)d\Lambda_j^a(s)}{\mathbb{P}_n\psi_2(s;a)}\right\} \\
&= E\left\{\int_0^t d\Lambda_j^a(s)\right\} = \Lambda_j^a(t).
\end{align*}
The fourth equation is because that $\overline{M}_j(t;a)$ is a martingale.

To calculate the finite-sample variance, we first note that
\begin{align*}
\var\{d\psi_{1j}(s;a)\mid\mathcal{F}_j(s;a)\} &= \var\left\{\frac{I(A=a)}{e(a;X)}dN_j(s)\mid\mathcal{F}_j(s;a)\right\} \\
&= \frac{I(A=a)}{e(a;X)^2} \var\{dN_j(s)\mid\mathcal{F}_j(s;a)\} \\
&= \frac{I(A=a)}{e(a;X)^2} Y(s)d\Lambda_j^a(s) \\
&= w(a;A,X) \psi_2(s;a)d\Lambda_j^a(s)
\end{align*}
by the property of martingales. Then
\begin{align*}
\var\{\widetilde\Lambda_j^a(t)\} &= \var\left\{\int_0^t\frac{\mathbb{P}_nd\psi_{1j}(s;a)}{\mathbb{P}_n\psi_2(s;a)}\right\} \\
&= \var\left\{\int_0^t\frac{\mathbb{P}_nd\psi_{1j}(s;a)-\mathbb{P}_n\psi_2(s;a)d\Lambda_j^a(s)}{\mathbb{P}_n\psi_2(s;a)}\right\} \\
&= \var\left\{\int_0^t\frac{d\overline{M}_j(s;a)}{\mathbb{P}_n\psi_2(s;a)}\right\} \\
&= E\left[\int_0^t\frac{\var\{\mathbb{P}_nd\psi_{1j}(s;a)\mid\mathcal{F}_j(s;a)\}}{\{\mathbb{P}_n\psi_2(s;a)\}^2}\right] \\
&= \frac{1}{n}E\left[\int_0^t\frac{\mathbb{P}_n\{w(a;A,X)\psi_2(s;a)\}d\Lambda_j^a(s)}{\{\mathbb{P}_n\psi_2(s;a)\}^2}\right].
\end{align*}

Next, we derive the influence function (IF) of $\widetilde\Lambda_j^a(t)$. Since
\[
\Lambda_j^a(t) = \int_0^t \frac{d\Psi_{1j}(s;a)}{\Psi_2(s;a)}
\]
is Hadamard derivable with respect to $\Psi_{1j}(t;a)$ and $\Psi_2(t;a)$, we can apply the functional delta method,
\begin{align*}
\IF\{\widetilde\Lambda_j^a(t)\} &= \int_0^t \frac{d\IF\{\widetilde\Psi_{1j}(s;a)\}\Psi_2(s;a)-d\Psi_{1j}(s;a)\IF\{\widetilde\Psi_2(s;a)\}}{\Psi_2(s;a)^2} \\
&= \int_0^t \frac{\psi_{1j}(s;a)\Psi_2(s;a)-d\Psi_{1j}(s;a)\psi_2(s;a)}{\Psi_2(s;a)^2} \\
&= \int_0^t \frac{\psi_{1j}(s;a)-\psi_2(s;a)d\Lambda_j^a(s)}{\Psi_2(s;a)} \\
&= \int_0^t \frac{1}{\Psi_2(s;a)}dM_j(s;a).
\end{align*}
That is,
\[
\widetilde\Lambda_j^a(t)-\Lambda_j^a(t) = \mathbb{P}_n\left[\IF\{\widetilde\Lambda_j^a(t)\}\right] + o_p(n^{-1/2}).
\]
By the central limit theory (CLT),
\[
\sqrt{n}\{\widetilde\Lambda_j^a(t)-\Lambda_j^a(t)\} \xrightarrow{d} N\left(0, E[\IF\{\widetilde\Lambda_j^a(t)\}^2]\right).
\]

According to the martingale central limit theorem, $\widetilde\Lambda_j^a(\cdot)-\Lambda_j^a(\cdot)$ weakly converges to a Gaussian process $G_j^a(\cdot)$ whose variance is $E[\IF\{\widetilde\Lambda_j^a(\cdot)\}^2]$. Furthermore, we can show the uniform convergence of $\widetilde\Lambda_j^a(\cdot)$.
By the inequality of Lenglart \citep{kalbfleisch2011statistical}, for any positive $\eta$ and $\delta$, we have
\begin{align*}
P\left\{\sup_{t \in [0,t^*]}|G_j^a(t)|> \sqrt{\eta} \right\} 
&\leq \frac{\delta}{\eta} + P\left[\int_0^{t^*} \frac{1}{\{\mathbb{P}_n\psi_2(s;a)\}^2} d\left<\overline{M}_j\right>(s;a) > \delta\right]\\
&\leq \frac{\delta}{\eta} + P\left[\int_0^{t^*} \frac{\mathbb{P}_n\{w(a;A,X)^2Y(s)\}d\Lambda_j^a(s)}{n\{\mathbb{P}_n\psi_2(s;a)\}^2} > \delta\right]\\
&\leq \frac{\delta}{\eta} + P\left[\int_0^{t^*} \frac{d\Lambda_j^a(s)}{nc^2\{\mathbb{P}_n\psi_2(t^*;a)\}^2} > \delta\right]\\
&\leq \frac{\delta}{\eta} + P\left[\frac{1}{nc^2\{\mathbb{P}_n\psi_2(t^*;a)\}^2} \Lambda_j^a(s) > \delta\right]\\
&\leq \frac{\delta}{\eta} + P\left[\mathbb{P}_n\psi_2(t^*;a) < \frac{\Lambda_j^{a}(t^*)}{nc^2\delta} \right] \\
&\to \frac{\delta}{\eta}
\end{align*}
as $n\to\infty$ because $\mathbb{P}_n\psi_2(t^*;a) = \Psi_2(t^*;a)+o_p(n^{-1/2})$ and $\Psi_2(t^*;a)>c$ by positivity. Since $\eta$ and $\delta$ are arbitrary, it follows that
\[
\sup_{t\in[0,t^*]} |\widetilde\Lambda_j^a(t)-\Lambda_j^a(t)| \xrightarrow{p} 0.
\]

\subsection{Theorem 3 (IF with estimated propensity score)}

We first consider $\Psi_{1j}(s;a)$. Let $\widehat\Psi_{1j}(s;a)$ be $\Psi_{1j}(s;a)$ with the propensity score substituted by its estimate. We decompose
\begin{align*}
\widehat\Psi_{1j}(s;a) - \Psi_{1j}(s;a) &= \mathbb{P}_n\widehat\psi_{1j}(s;a) - \mathbb{P}\psi_{1j}(s;a) \\
&= (\mathbb{P}_n-\mathbb{P})\psi_{1j}(s;a) + (\mathbb{P}_n-\mathbb{P})(\widehat\psi_{1j}-\psi_{1j})(s;a) + \mathbb{P}(\widehat\psi_{1j}-\psi_{1j})(s;a) \\
&= R_1 + R_2 + R_3.
\end{align*}
The first term 
\[
R_1 = (\mathbb{P}_n-\mathbb{P})\psi_{1j}(s;a) = \mathbb{P}_n(\psi_{1j}-\Psi_{1j})(s;a).
\]
The second term 
\[
R_2 = (\mathbb{P}_n-\mathbb{P})(\widehat\psi_{1j}-\psi_{1j})(s;a) = o_p(n^{-1/2})
\]
because $\psi_{1j}(s;a)$, a weighted indicator function by inverse of propensity score (Donsker), belongs to a Donsker class \citep{wellner2013weak}.
The third term 
\[
R_3 = \mathbb{P}(\widehat\psi_{1j}-\psi_{1j})(s;a)
\]
represents the additional variation due to the estimated propensity score. 

Assume that $\widehat\theta$ in the propensity score is regularly asymptotically linear (RAL), 
\[
\widehat\theta - \theta = \mathbb{P}_n\phi + o_p(n^{-1/2}).
\]
Then
\begin{align*}
R_3 &= \mathbb{P}(\widehat\psi_{1j}-\psi_{1j})(s;a) \\
&= \mathbb{P}\left[I(A=a, T\leq s, \Delta=j)\left\{\frac{1}{e(a;X;\widehat\theta)}-\frac{1}{e(a;X;\theta)}\right\}\right] \\
&= - \mathbb{P}\left[I(A=a, T\leq s, \Delta=j) \left\{\frac{\dot{e}(a;X;\theta)}{e(a;X;\theta)^2}(\widehat\theta-\theta) + o_p(\|\widehat\theta-\theta\|)\right\}\right] \\
&= - \mathbb{P}\left\{I(A=a, T\leq s, \Delta=j) \frac{\dot{e}(a;X;\theta)}{e(a;X;\theta)^2}\right\} \mathbb{P}_n\phi + o_p(n^{-1/2}) \\
&= B_{1j}(s;a) \mathbb{P}_n\phi + o_p(n^{-1/2}),
\end{align*}
where 
\[
B_{1j}(s;a) = - \mathbb{P}\left\{I(A=a, T\leq s, \Delta=j) \frac{\dot{e}(a;X;\theta)}{e(a;X;\theta)^2}\right\}.
\]
Therefore,
\begin{align*}
\widehat\Psi_{1j}(s;a)-\Psi_{1j}(s;a) &= \mathbb{P}_n(\psi_{1j}-\Psi_{1j})(s;a) + B_{1j}(s;a)\mathbb{P}_n\phi + o_p(n^{-1/2}).
\end{align*}

In the same manner,
\begin{align*}
\widehat\Psi_{2}(s;a)-\Psi_{2}(s;a) &= \mathbb{P}_n(\psi_{2}-\Psi_{2})(s;a) + B_{2}(s;a)\mathbb{P}_n\phi + o_p(n^{-1/2}),
\end{align*}
where
\[
B_{2}(s;a) = - \mathbb{P}\left\{I(A=a, T\geq s) \frac{\dot{e}(a;X;\theta)}{e(a;X;\theta)^2}\right\}.
\]
So we have the influence functions for $\widehat\Psi_{1j}(s;a)$ and $\widehat\Psi_2(s;a)$,
\begin{align*}
\IF\{\widehat\Psi_{1j}(s;a)\} &= \psi_{1j}(s;a) - \Psi_1(s;a) + B_{1j}(s;a)\phi, \\
\IF\{\widehat\Psi_{2}(s;a)\} &= \psi_2(s;a) - \Psi_2(s;a) + B_2(s;a)\phi.
\end{align*}

Since $\Lambda_j^a(t)$ is Hadamard derivable with respect to $\Psi_{1j}(t;a)$ and $\Psi_2(t;a)$, so we can apply functional delta method,
\begin{align*}
\IF\{\widehat\Lambda_j^a(t)\} &= \int_0^t \frac{d\IF\{\widehat\Psi_{1j}(s;a)\}\Psi_2(s;a)-d\Psi_{1j}(s;a)\IF\{\widehat\Psi_2(s;a)\}}{\Psi_2(s;a)^2} \\
&= \int_0^t \frac{1}{\Psi_2(s;a)}\left[d\IF\{\widehat\Psi_{1j}(s;a)\}-\IF\{\widehat\Psi_{2}(s;a)\}d\Lambda_j^a(s)\right] \\
&= \int_0^t \frac{1}{\Psi_2(s;a)}\left[dM_j(s;a)+(dB_{1j}-B_2)(s;a)d\Lambda_j^a(s)\phi\right]. 
\end{align*}
Furthermore, we can simplify the augmented term
\begin{align*}
\nu_j(t;a) &= \int_0^t \frac{1}{\Psi_2(s;a)}(dB_{1j}-B_2)(s;a)d\Lambda_j^a(s) \\
&= - \int_0^t \frac{1}{\Psi_2(s;a)} \mathbb{P}\left\{dM_j(s;a)\frac{\dot{e}(a,X;\theta)}{e(a,X;\theta)^2}\right\}.
\end{align*}
Since $\overline{M}_j(s;a)$ is not a martingale with respect to the filter $\overline{\mathcal{F}}_j(t;a) = \{Y_i(s), w(a;A_i,X_i), X_i: s\leq t, i=1,\ldots,n\}$, there is no guarantee that $\nu_j(t;a)=0$.

Finally, we have
\[
\widehat\Lambda_j^a(t)-\Lambda_j^a(t) = \mathbb{P}_n\left[\IF\{\widehat\Lambda_j^a(t)\}\right] + o_p(n^{-1/2}).
\]
By the central limit theory, 
\[
\sqrt{n}\{\widehat\Lambda_j^a(t)-\Lambda_j^a(t)\} \xrightarrow{d} N\left(0, E[\IF\{\widehat\Lambda_j^a(t)\}^2]\right).
\]

In fact, $\widehat\Lambda_j^a(\cdot)-\Lambda_j^a(\cdot)$ weakly converges to a Gaussian process whose variance is $E[\IF\{\widehat\Lambda_j^a(\cdot)\}^2]$ and
\[
\sup_{t\in[0,t^*]} |\widehat\Lambda_j^a(t)-\Lambda_j^a(t)| \xrightarrow{p} 0.
\]
The proof of uniform convergence is similar to that in the proceeding section, as long as $c<\widehat{e}(a;X)<1-c$ with probability 1.

\subsection{From hazards to incidences}

Hereafter we assume the cause-specific hazard $\Lambda_j^a(t)$ is estimated by $\widehat\Lambda_j^a(t)$. The cumulative incidence function (CIF) of event $j$
\[
F_j^a(t) = \int_0^t \exp\left\{-\sum_{k=1}^{J}\Lambda_k^a(s)\right\} d\Lambda_j^a(s)
\]
is Hadamard derivable with respect to the cause-specific hazards $\{\Lambda_k^a(t): k=1,\ldots,J\}$. So we apply the functional delta method,
\begin{align*}
\IF\{\widehat{F}_j^a(t)\} &= \int_0^t \exp\left\{-\sum_{k=1}^{J}\Lambda_k^a(s)\right\} d\IF\{\widehat\Lambda_j^a(s)\} \\
&\quad - \int_0^t \exp\left\{-\sum_{k=1}^{J}\Lambda_k^a(s)\right\}\sum_{k=1}^{J}\IF\{\widehat\Lambda_k^a(s)\}d\Lambda_j^a(s) \\
&= \int_0^t \overline{F}^a(s)d\IF\{\widehat\Lambda_j^a(s)\} - \sum_{k=1}^{J}\int_0^t \IF\{\widehat\Lambda_k^a(s)\}dF_j^a(s),
\end{align*}
where $\overline{F}^a(s) = 1-\sum_{k=1}^{J}F_k^a(s)$ is the overall survival function. By the central limit theory,
\[
\sqrt{n}\{\widehat{F}_j^a(t)-F_j^a(t)\} \xrightarrow{d} N\left(0, E[\IF\{\widehat{F}_j^a(t)\}^2]\right).
\]
In fact, $\widehat{F}_j^a(\cdot)-F_j^a(\cdot)$ weakly converges to a Gaussian process with variance $E[\IF\{\widehat{F}_j^a(\cdot)\}^2]$.

\section{Efficient influence function}

The efficient influence function (EIF) of $\Lambda_j^a(t;x)$ is \citep{martinussen2023estimation, rytgaard2024targeted}
\[
\EIF\{\Lambda_j^a(t;x)\} = \frac{I\{A=a\}}{p(x)P(A=a\mid X)} \int_0^t \frac{dM_j(s;A,X)}{P(T>t\mid A,X)},
\]
where
\[
M_j(t;a,x) = I\{A=a, X=x\} \int_0^t \{dN_j(s) - Y(s)d\Lambda_j^a(s;x)\}
\]
and
\[
P(T>t\mid A=a,X) = P(\tilde{T}^a>t\mid X) P(\tilde{C}^a>t\mid X).
\]
We assume random censoring given covariates (generally, this is weaker than completely random censoring). Let $\Lambda_0^a(t;x)$ be the hazard of censoring, then
\[
P(T>t \mid A=a,X) = \exp\left\{-\sum_{k=0}^{J}\Lambda_k^a(t;X)\right\}.
\]
Thus, an influence function of $F_j^a(t)$ is given by
\begin{align*}
\IF\{F_j^a(t)\} &= \frac{I\{A=a\}}{P(A=a\mid X)} \int_0^t \exp\left\{-\sum_{k=1}^{J}\Lambda_k^a(s;X)\right\} \\
&\quad \left[\frac{dM_j(s;A,X)}{P(T>s\mid A,X)}-\sum_{k=1}^{J}\int_0^s\frac{dM_k(u;A,X)}{P(T>s\mid A,X)} d\Lambda_j^a(s;X)\right] + F_j^a(t;X) - F_j^a(t).
\end{align*}
We can show that this influence function is in the tangent space (by integration by part)
\[
\dot{\mathcal{T}} = \bigoplus_{k=0}^{J} \dot{\mathcal{T}}_k \oplus \dot{\mathcal{P}},
\]
where
\begin{align*}
\dot{\mathcal{T}}_k &= \left\{\int_0^{t^*}h(s,A,X)dM_k(s;A,X)\right\}, \\
\dot{\mathcal{P}} &= \left\{h(A,X): E[h(A,X)]=0\right\}.
\end{align*}
Therefore, this influence function is the efficient influence function.

By plugging in the fitted models and solving the estimating equation that the empirical EIF equals zero, we obtain the EIF-based estimator of $F_j^a(t)$. This estimator does not have the Nelson--Aalen form. Interestingly, this estimator has semiparametric efficiency and multiple robustness under weak conditions. The estimator is consistent if (1) all cause-specific hazards are correctly specified, or (2) the propensity score and censoring probability model are correctly specified and all but one cause-specific hazards are correctly specified. In practice, the propensity score is fitted at baseline whereas the censoring probability and cause-specific hazards are fitted using post-treatment data. Modeling the univariate propensity score is computationally easier by regression and provides desirable asymptotic properties. It is challenging to correctly specify the hazards if the dependence of events is complex. Misspecification of working models may lead to bias and inconsistent variance estimation.

\section{Additional simulation results}

Table \ref{tab:SF1_2000} and Table \ref{tab:SF0_2000} show the estimation results when the sample size $n=2000$. With a larger sample size, the bias is smaller. The coverage rate of the confidence interval is close to 95\% using the corrected standard error for the adjusted Nelson--Aalen estimator. Still, the coverage rate of the confidence interval using the naive standard error is slightly lower.

\begin{table}[!tbh]
\centering
\caption{Bias, standard deviation, mean standard error and confidence interval coverage rate of some estimators for the counterfactual cumulative incidence function $F_1^1(t)$ when the sample size $n=2000$} \label{tab:SF1_2000}
\begin{tabular}{lcccccccc}
  \toprule
Time & 1 & 2 & 3 & 4 & 5 & 6 & 7 & 8 \\ 
  \midrule
\multicolumn{9}{l}{Bias} \\
Oracle & -0.000 & -0.000 & -0.001 & -0.000 & -0.000 & -0.000 & -0.000 & -0.001 \\ 
Adjusted NA & -0.000 & -0.000 & -0.000 & -0.000 & -0.000 & -0.000 & -0.000 & -0.001 \\ 
Weighted AJ & -0.000 & -0.000 & -0.000 & 0.000 & 0.000 & 0.000 & 0.000 & 0.000 \\ 
  \midrule
\multicolumn{9}{l}{Standard deviation (SD)} \\
Oracle & 0.005 & 0.009 & 0.012 & 0.014 & 0.015 & 0.016 & 0.016 & 0.017 \\ 
Adjusted NA & 0.005 & 0.009 & 0.012 & 0.014 & 0.015 & 0.016 & 0.016 & 0.017 \\ 
Weighted AJ & 0.005 & 0.009 & 0.012 & 0.014 & 0.015 & 0.016 & 0.016 & 0.017 \\ 
  \midrule
\multicolumn{9}{l}{Standard error (SE)} \\
Oracle & 0.005 & 0.009 & 0.012 & 0.014 & 0.015 & 0.016 & 0.016 & 0.016 \\ 
Naive & 0.005 & 0.009 & 0.012 & 0.014 & 0.015 & 0.016 & 0.016 & 0.016 \\ 
Corrected & 0.005 & 0.009 & 0.012 & 0.014 & 0.016 & 0.016 & 0.016 & 0.017 \\ 
Bootstrap & 0.005 & 0.009 & 0.012 & 0.014 & 0.015 & 0.015 & 0.016 & 0.016 \\ 
Weighted AJ & 0.005 & 0.009 & 0.012 & 0.014 & 0.015 & 0.016 & 0.016 & 0.016 \\ 
  \midrule
\multicolumn{9}{l}{Coverage rate} \\
Oracle & 0.942 & 0.943 & 0.933 & 0.944 & 0.946 & 0.943 & 0.945 & 0.954 \\ 
Naive & 0.941 & 0.950 & 0.934 & 0.945 & 0.946 & 0.950 & 0.949 & 0.953 \\ 
Corrected & 0.943 & 0.954 & 0.943 & 0.952 & 0.950 & 0.957 & 0.952 & 0.960 \\ 
Bootstrap & 0.934 & 0.944 & 0.932 & 0.941 & 0.943 & 0.941 & 0.947 & 0.947 \\ 
Weighted AJ & 0.942 & 0.951 & 0.936 & 0.946 & 0.947 & 0.949 & 0.949 & 0.953 \\ 
\bottomrule
\end{tabular}
\end{table}

\begin{table}[!tbh]
\centering
\caption{Bias, standard deviation, mean standard error and confidence interval coverage rate of some estimators for the counterfactual cumulative incidence function $F_1^0(t)$ when the sample size $n=2000$} \label{tab:SF0_2000}
\begin{tabular}{lcccccccc}
  \toprule
Time & 1 & 2 & 3 & 4 & 5 & 6 & 7 & 8 \\ 
  \midrule
\multicolumn{9}{l}{Bias} \\
Oracle & 0.000 & 0.000 & 0.000 & 0.000 & 0.000 & -0.000 & -0.000 & -0.000 \\ 
Adjusted NA & 0.000 & 0.000 & 0.000 & 0.000 & -0.000 & -0.000 & -0.001 & -0.000 \\ 
Weighted AJ & 0.000 & 0.001 & 0.001 & 0.001 & 0.001 & 0.001 & 0.001 & 0.001 \\ 
  \midrule
\multicolumn{9}{l}{Standard deviation (SD)} \\
Oracle & 0.013 & 0.015 & 0.016 & 0.017 & 0.018 & 0.018 & 0.018 & 0.018 \\ 
Adjusted NA & 0.013 & 0.015 & 0.016 & 0.017 & 0.018 & 0.018 & 0.018 & 0.018 \\ 
Weighted AJ & 0.013 & 0.015 & 0.016 & 0.017 & 0.018 & 0.018 & 0.018 & 0.018 \\ 
  \midrule
\multicolumn{9}{l}{Standard error (SE)} \\
Oracle & 0.012 & 0.015 & 0.017 & 0.017 & 0.017 & 0.018 & 0.018 & 0.018 \\ 
Naive & 0.012 & 0.015 & 0.017 & 0.017 & 0.018 & 0.018 & 0.018 & 0.018 \\ 
Corrected & 0.013 & 0.015 & 0.017 & 0.017 & 0.018 & 0.018 & 0.018 & 0.018 \\ 
Bootstrap & 0.012 & 0.015 & 0.016 & 0.017 & 0.017 & 0.017 & 0.018 & 0.018 \\ 
Weighted AJ & 0.013 & 0.015 & 0.017 & 0.017 & 0.018 & 0.018 & 0.018 & 0.018 \\ 
  \midrule
\multicolumn{9}{l}{Coverage rate} \\
Oracle & 0.940 & 0.949 & 0.957 & 0.948 & 0.940 & 0.943 & 0.938 & 0.941 \\ 
Naive & 0.935 & 0.954 & 0.960 & 0.952 & 0.946 & 0.945 & 0.940 & 0.942 \\ 
Corrected & 0.938 & 0.954 & 0.960 & 0.952 & 0.946 & 0.948 & 0.945 & 0.945 \\ 
Bootstrap & 0.929 & 0.953 & 0.954 & 0.944 & 0.943 & 0.942 & 0.944 & 0.942 \\ 
Weighted AJ & 0.934 & 0.953 & 0.957 & 0.949 & 0.946 & 0.945 & 0.945 & 0.945 \\ 
\bottomrule
\end{tabular}
\end{table}

To conduct a sensitivity analysis, now we suppose that the propensity score is misspecified. We consider two misspecified propensity scores: (1) the constant propensity score $e^{\star}(1;x)=\theta$, and (2) the probit propensity score $e^*(1;x) = \Phi(\theta_0 + x'\theta_1)$ where $\Phi(\cdot)$ is the distribution function of the standard normal distribution. The parameters in the propensity score can be estimated from data. Of note, if a constant propensity score is used, then $\dot{e}(a;X;\theta)/e(a;X;\theta)^2$ is a constant, and thus $\nu_j(t;a)=0$. The corrected standard error would be identical to the naive standard error.

Table \ref{tab:S1_500} and Table \ref{tab:S0_500} show the bias, standard deviation, standard error and coverage rate of nominal 95\% confidence intervals when the propensity score model is misspecified and the sample size $n=500$. In this setting, the bias is still small, but the coverage rate of the confidence interval may slightly deviate from 95\%. We also find that estimating the propensity score by probit regression leads to a larger standard error than logistic regression or constant. This is because the probit function has a faster descent rate at the tail, so $\dot{e}(a;X;\theta)/e(a;X;\theta)^2$ has a larger variation.

\begin{table}[!tbh]
\centering
\caption{Estimation for $F_1^1(t)$ when the propensity score is misspecified and $n=500$} \label{tab:S1_500}
\begin{tabular}{lcccccccc}
  \toprule
Time & 1 & 2 & 3 & 4 & 5 & 6 & 7 & 8 \\ 
  \midrule
  \multicolumn{9}{l}{Bias} \\
Constant & -0.001 & -0.003 & -0.006 & -0.007 & -0.008 & -0.007 & -0.007 & -0.006 \\ 
Probit & -0.001 & -0.002 & -0.003 & -0.003 & -0.003 & -0.004 & -0.005 & -0.005 \\ 
  \midrule
  \multicolumn{9}{l}{Standard deviation (SD)} \\
Constant & 0.008 & 0.016 & 0.022 & 0.026 & 0.028 & 0.028 & 0.029 & 0.030 \\  
Probit & 0.025 & 0.031 & 0.034 & 0.034 & 0.035 & 0.036 & 0.037 & 0.037 \\ 
  \midrule
  \multicolumn{9}{l}{Standard error (SE)} \\
Constant (Naive) & 0.008 & 0.016 & 0.021 & 0.025 & 0.028 & 0.029 & 0.030 & 0.030 \\ 
Constant (Corrected) & 0.008 & 0.016 & 0.021 & 0.025 & 0.028 & 0.029 & 0.030 & 0.030 \\ 
Constant (Bootstrap) & 0.008 & 0.016 & 0.021 & 0.025 & 0.028 & 0.029 & 0.029 & 0.030 \\
Probit (Naive) & 0.025 & 0.030 & 0.033 & 0.034 & 0.035 & 0.035 & 0.035 & 0.035 \\ 
Probit (Corrected) & 0.038 & 0.052 & 0.060 & 0.066 & 0.071 & 0.074 & 0.075 & 0.077 \\ 
Probit (Bootstrap) & 0.024 & 0.030 & 0.032 & 0.033 & 0.034 & 0.034 & 0.034 & 0.035 \\ 
  \midrule
  \multicolumn{9}{l}{Coverage rate} \\
Constant (Naive) & 0.883 & 0.923 & 0.919 & 0.915 & 0.936 & 0.952 & 0.957 & 0.949 \\
Constant (Corrected) & 0.883 & 0.923 & 0.919 & 0.915 & 0.936 & 0.952 & 0.957 & 0.949 \\
Constant (Bootstrap) & 0.883 & 0.921 & 0.914 & 0.917 & 0.935 & 0.953 & 0.947 & 0.947 \\
Probit (Naive) & 0.934 & 0.943 & 0.934 & 0.942 & 0.946 & 0.942 & 0.934 & 0.938 \\ 
Probit (Corrected) & 0.965 & 0.980 & 0.976 & 0.977 & 0.979 & 0.984 & 0.981 & 0.981 \\ 
Probit (Bootstrap) & 0.933 & 0.939 & 0.927 & 0.937 & 0.938 & 0.938 & 0.936 & 0.935 \\
   \bottomrule
\end{tabular}
\end{table}

\begin{table}[!tbh]
\centering
\caption{Estimation for $F_1^0(t)$ when the propensity score is misspecified and $n=500$} \label{tab:S0_500}
\begin{tabular}{lcccccccc}
  \toprule
Time & 1 & 2 & 3 & 4 & 5 & 6 & 7 & 8 \\ 
  \midrule
  \multicolumn{9}{l}{Bias} \\
Constant & 0.004 & 0.003 & 0.000 & -0.002 & -0.005 & -0.008 & -0.011 & -0.013 \\ 
Probit & -0.000 & -0.001 & -0.001 & -0.000 & -0.000 & -0.001 & -0.002 & -0.003 \\ 
  \midrule
  \multicolumn{9}{l}{Standard deviation (SD)} \\
Constant & 0.023 & 0.029 & 0.031 & 0.030 & 0.032 & 0.032 & 0.033 & 0.033 \\ 
Probit & 0.009 & 0.017 & 0.024 & 0.028 & 0.029 & 0.029 & 0.030 & 0.032 \\  
  \midrule
  \multicolumn{9}{l}{Standard error (SE)} \\
Constant (Naive) & 0.024 & 0.028 & 0.031 & 0.032 & 0.032 & 0.032 & 0.032 & 0.033 \\ 
Constant (Corrected) & 0.024 & 0.028 & 0.031 & 0.032 & 0.032 & 0.032 & 0.032 & 0.033 \\ 
Constant (Bootstrap) & 0.023 & 0.028 & 0.030 & 0.031 & 0.032 & 0.032 & 0.032 & 0.032 \\ 
Probit (Naive) & 0.009 & 0.017 & 0.023 & 0.028 & 0.030 & 0.031 & 0.032 & 0.032 \\ 
Probit (Corrected) & 0.010 & 0.020 & 0.029 & 0.033 & 0.036 & 0.037 & 0.037 & 0.037 \\ 
Probit (Bootstrap) & 0.009 & 0.017 & 0.023 & 0.027 & 0.029 & 0.030 & 0.031 & 0.031 \\
  \midrule
  \multicolumn{9}{l}{Coverage rate} \\
Constant (Naive) & 0.955 & 0.946 & 0.943 & 0.954 & 0.945 & 0.938 & 0.928 & 0.929 \\ 
Constant (Corrected) & 0.955 & 0.946 & 0.943 & 0.954 & 0.945 & 0.938 & 0.928 & 0.929 \\ 
Constant (Bootstrap) & 0.953 & 0.944 & 0.945 & 0.951 & 0.941 & 0.938 & 0.931 & 0.929 \\
Probit (Naive) & 0.888 & 0.931 & 0.945 & 0.947 & 0.955 & 0.957 & 0.965 & 0.952 \\
Probit (Corrected) & 0.889 & 0.940 & 0.958 & 0.959 & 0.966 & 0.971 & 0.972 & 0.969 \\ 
Probit (Bootstrap) & 0.885 & 0.930 & 0.941 & 0.929 & 0.945 & 0.954 & 0.960 & 0.946 \\
   \bottomrule
\end{tabular}
\end{table}

\bibliographystyle{apalike}
\bibliography{ref}